# FM Backscatter: Enabling Connected Cities and Smart Fabrics

*Anran Wang*[†], *Vikram Iyer*[†], *Vamsi Talla, Joshua R. Smith and Shyamnath Gollakota*
*University of Washington*
[†]Co-primary Student Authors

**Abstract** – This paper enables connectivity on everyday objects by transforming them into FM radio stations. To do this, we show for the first time that ambient FM radio signals can be used as a signal source for backscatter communication. Our design creates backscatter transmissions that can be decoded on any FM receiver including those in cars and smartphones. This enables us to achieve a previously infeasible capability: backscattering information to cars and smartphones in outdoor environments.

Our key innovation is a modulation technique that transforms backscatter, which is a multiplication operation on RF signals, into an addition operation on the audio signals output by FM receivers. This enables us to embed both digital data as well as arbitrary audio into ambient analog FM radio signals. We build prototype hardware of our design and successfully embed audio transmissions over ambient FM signals. Further, we achieve data rates of up to 3.2 kbps and ranges of 5–60 feet, while consuming as little as 11.07 $\mu$W of power. To demonstrate the potential of our design, we also fabricate our prototype on a cotton t-shirt by machine sewing patterns of a conductive thread to create a smart fabric that can transmit data to a smartphone. We also embed FM antennas into posters and billboards and show that they can communicate with FM receivers in cars and smartphones.

## 1 Introduction

This paper asks the following question: can we enable everyday objects in outdoor environments to communicate with cars and smartphones, without worrying about power? Such a capability can enable transformative visions such as connected cities and smart fabrics that promise to change the way we interact with objects around us. For instance, bus stop posters and street signs could broadcast digital content about local attractions or advertisements directly to a user's car or smartphone. Posters advertising local artists could broadcast clips of their music or links to purchase tickets for upcoming shows. A street sign could broadcast information about the name of an intersection, or when it is safe to cross a street to improve accessibility for the disabled. Such ubiquitous low-power connectivity would also enable smart fabric applications — smart clothing with sensors integrated into the fabric itself could monitor a runner's gait or vital signs and directly transmit the information to their phone.

While recent efforts on backscatter communication [40, 49, 38, 36] dramatically reduce the power requirements for wireless transmissions, they are unsuitable for outdoor environments. Specifically, existing approaches either use custom transmissions from RFID readers or backscatter ambient Wi-Fi [37] and TV transmissions [40, 42]. RFID-based approaches are expensive in outdoor environments given the cost of deploying the reader infrastructure. Likewise, while Wi-Fi backscatter [37] is useful indoors, it is unsuitable for outdoor environments. Finally, TV signals are available outdoors, but smartphones as well as most cars do not have TV receivers and hence cannot decode the backscattered signals.

Taking a step back, the requirements for our target applications are four-fold: 1) The ambient signals we hope to backscatter must be ubiquitous in outdoor environments, 2) devices such as smartphones and cars must have the receiver hardware to decode our target ambient signals, 3) it should be legal to backscatter in the desired frequencies without a license, which precludes cellular transmissions [31], and 4) in order to retrieve the backscattered data, we should ideally have a software-defined radio like capability to process the raw incoming signal without any additional hardware.

Our key contribution is the observation that FM radio satisfies the above constraints. Broadcast FM radio infrastructure already exists in cities around the world. These FM radio towers transmit at a high power of several hundred kilowatts [31] which provides an ambient signal source that can be used for backscatter communication. Additionally, FM radio receivers are included in the LTE and Wi-Fi chipsets of almost every smartphone [24] and have recently been activated on Android devices in the United States [23, 8]. Further, the FCC provides an exemption for low-power transmitters to operate on FM bands without requiring a license [30]. Finally, unlike commercial Wi-Fi, Bluetooth and cellular chipsets that provide only packet level access, FM radios provide access to the raw audio decoded by the receiver. These raw audio signals can be used in lieu of a software-defined radio to extract the backscattered data.

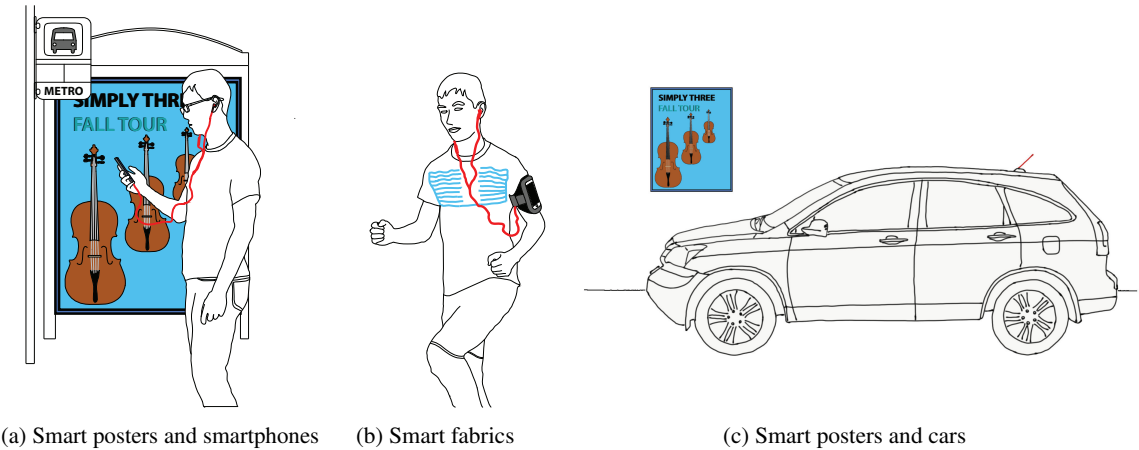

(a) Smart posters and smartphones  (b) Smart fabrics  (c) Smart posters and cars

Figure 1: **Example applications enabled by backscattering FM radio signals.** (a) smart posters capable of broadcasting music to nearby users; (b) vital signs monitoring with sensors integrated into smart fabrics; (c) FM receivers in cars receiving music sent from nearby smart posters.

Building on this, we transform everyday objects into FM radio stations. Specifically, we design the first system that uses FM signals as an RF source for backscatter. We show that the resulting transmissions can be decoded on any FM receiver including those in cars and smartphones. Achieving this is challenging for two key reasons:

• Unlike software radios that give raw RF samples, FM receivers output only the demodulated audio. This complicates decoding since the backscatter operation is performed on the RF signals corresponding to the FM transmissions while an FM receiver outputs only the demodulated audio. Thus, the backscatter operation has to be designed to be compatible with the FM demodulator.

• FM stations broadcast audio that ranges from news channels with predominantly human speech, to music channels with a richer set of audio frequencies. In addition, they can broadcast either a single stream (mono mode) or two different audio streams (stereo mode) to play on the left and right speakers. Ideally, the backscatter modulation should operate efficiently with all these FM modes and audio characteristics.

To address these challenges, we leverage the structure of FM radio signals. At a high level, we introduce a modulation technique that transforms backscatter, which is a multiplication operation on RF signals, into an addition operation on the audio signal output by FM receivers (see §3.3). This allows us to embed audio and data information in the underlying FM audio signals. Specifically, we use backscatter to synthesize baseband transmissions that imitate the structure of FM signals, which makes it compatible with the FM demodulator. Building on this, we demonstrate three key capabilities.

• *Overlay backscatter.* We overlay arbitrary audio on ambient FM signals to create a composite audio signal that can be heard using any FM receiver, without any additional processing. We also design a modulation technique that overlays digital data which can be decoded on FM receivers with processing capabilities, e.g., smartphones.

• *Stereo backscatter.* A number of FM stations, while operating in the stereo mode, do not effectively utilize the stereo stream. We backscatter data and audio on these under-utilized stereo streams to achieve a low interference communication link. Taking it a step further, we can also trick FM receivers to decode mono FM signals in the stereo mode, by inserting the pilot signal that indicates a stereo transmission. This allows us to backscatter information in the interference-free stereo stream.

• *Cooperative backscatter.* Finally, we show that using cooperation between two smartphones from users in the vicinity of the backscattering objects, we can imitate a MIMO system that cancels out the audio in the underlying ambient FM transmission. This allows us to decode the backscatter transmissions without any interference.

To evaluate our design, we first perform a survey of the FM signal strengths in a major metropolitan city and identify unused spectrum in the FM band, which we then use in our backscatter system. We implement a prototype FM backscatter device using off-the-shelf components and use it to backscatter data and arbitrary audio directly to a Moto G1 smartphone and 2010 Honda CRV's FM receiver. We compare the trade offs for the three backscat-ter techniques described above and achieve data rates of up to 3.2 kbps and ranges of 5–60 ft. Finally, we design and simulate an integrated circuit that backscatters audio signals, and show that it consumes only 11.07 $\mu$W.

To demonstrate the potential of our design, we build two proof-of-concept applications. We build and evaluate posters with different antenna topologies including dipoles and bowties fabricated using copper tape. We also fabricate our prototype on a cotton t-shirt by machine sewing patterns of a conductive thread and build a smart fabric that can transmit vital sign data to the phone, even when the user is walking or running.

**Contributions.** We use backscatter to transform everyday objects into FM radio stations. In particular, we

• Introduce the first system to backscatter audio and data to cars and smartphones, outdoors. We present the first backscatter design that uses FM radio as its signal source and creates signals that are decodable on any FM receiver.

• Transform backscatter, which is a multiplication operation on RF signals, into an addition operation on the FM audio signals. Building on this, we present three key capabilities: overlay, stereo and cooperative backscatter.

• Build FM backscatter hardware and evaluate it in different scenarios. To show the potential of our design, we also build two proof-of-concept application prototypes for smart fabrics and smart bus stop billboards.

## 2 Application Requirements

Current radio technologies cannot satisfy the requirements for the connected city and smart fabric applications described above. Traditional radio transmitters are relatively expensive to deploy at scale and consume significant amounts of power; in contrast, we develop a backscatter system that consumes orders of magnitude less power and can be produced at lower cost. In the remainder of this section we outline the requirements of our target applications in detail and explain why FM backscatter is the only viable approach.

*Connected cities.* Connected city devices such as posters and signs must continuously broadcast information to anyone passing by. Considering most outdoor locations such as signposts and bus stops do not have dedicated connections to power infrastructure, these devices will either be battery powered, or harvest power from the environment. A practical communication solution should seek to maximize battery life so as to prevent recurring maintenance costs for these objects. A normal FM transmitter designed for battery powered applications [15] consumes 18.8 mA when transmitting. If the chip is continuously broadcasting, this system would last less than 12 hrs using a 225 mAh battery coin cell battery [12]. In practice the battery life would likely be much shorter considering the current draw of the FM chip is significantly higher than the rated 0.2 mA discharge current with which the battery was tested. In contrast, our backscatter system could continuously transmit for almost 3 years. Additionally, even at scale an FM radio chip costs over $4 [16] whereas a backscatter solution costs as little as a few cents [9].

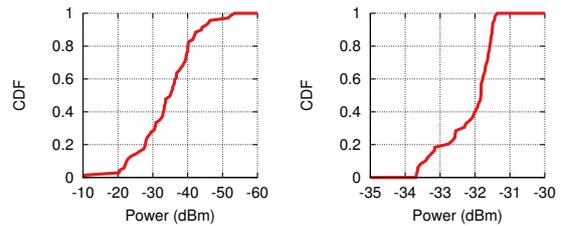

(a) Across a major US city.    (b) Across 24 hours of a day.

Figure 2: **Survey of FM radio signals.** (a) CDF of received power of FM radio signals across a major metropolitan city and (b) CDF of powers at a fixed location over a 24 hour duration.

Another potential alternative is Bluetooth Low Energy, such as the iBeacon standard that is designed for sending broadcast messages. In broadcast mode however, the BLE protocol is limited to sending short packets every 100 ms, and therefore not a viable solution for streaming audio. Additionally, while some cars have Bluetooth connectivity, their Bluetooth antennas are positioned inside to interact with mobile phones and other devices. An antenna inside the car would be shielded from our smart objects by the car's metal body, causing significant attenuation. In contrast, FM antennas are already positioned outside the car as this is where FM signals are strongest.

*Smart fabrics.* In addition to the same power constraints described above, smart fabrics require thin and flexible form factors. Materials such as conductive thread allow us to produce flexible FM antennas on textile substrates, which blend into cloth. However, commercially available flexible batteries have a limited discharge current. For example current flexible batteries on the market are limited to a 10 mA peak current [6] and cannot satisfy the requirements of FM or BLE radios when transmitting [15, 18]. While the reader may assume that the size of FM antennas limits smart fabric applications, we demonstrate that by using conductive thread we can take advantage of the whole surface area available on a garment to produce a flexible and seamlessly integrated antenna.

## 3 System Design

We use backscatter to encode audio and digital data on FM radio signals. In this section we begin by evaluating the availability and characteristics of FM radio broadcasts in urban areas. Next we provide background on FM radio and how we can leverage its signal structure for backscatter communication. Finally, we describe our encoding mechanism that embeds digital data using backscatter.

## 3.1 Survey of FM Radio Signals

Public and commercial FM radio broadcasts are a standard in urban centers around the world and provide a source of ambient RF signals in these environments. In order to cover a wide area and allow for operation despite complex multipath from structures and terrains, FM radio stations broadcast relatively high power signals. In the United States, FM radio stations produce an effective radiated power of up to 100 kW [31].

In this section, we survey the signal strength of FM radio transmissions across Seattle, WA. We drive through the city and measure the power of ambient FM radio signals using a software defined radio (SDR, USRP E310) connected to a quarter-wavelength monopole antenna (Diamond SRH789). Since there are multiple FM radio stations in most US cities, we record signals across the full 88–108 MHz FM spectrum and identify the FM station with the maximum power at each measurement location. We calibrate the raw SDR values to obtain power measurements in dBm using reference power measurements performed with a spectrum analyzer (Tektronix MDO4054B-3). We divide the surveyed area into 0.8 mi× 0.8 mi grid squares and determine the median power in each for a total of 69 measurements.

Fig. 2a shows the CDF of the measurements across the city. The plot shows that the FM signal strength varies between -10 and -55 dBm. We also note that the median signal strength across all the locations is -35.15 dBm, which is surprisingly high, despite measurements being performed in a dense urban environment with tall buildings and across elevation differences of 450 ft.[1] We note that, prior work on ambient backscatter reflecting TV signals, requires the strength of TV signals to be at least -10 to -20 dBm. In contrast, our goal is to decode the backscattered signals on FM receivers which have a sensitivity around -100 dBm [14, 1]. Thus, FM signals are promising as a signal source for backscatter.

Next, we place the same SDR setup in a fixed outdoor location over the course of 24 hours. We record 100,000 samples over the whole FM radio band and compute the average power of the FM station with the highest power once every minute. Fig. 2b shows that the power varies with a standard deviation of 0.7 dBm, which shows that the received power is roughly constant across time. These measurements show that FM signals can provide a reliable signal source for backscatter.

---
[1]For comparison, at -50 dBm Wi-Fi can support hundreds of Mbps.

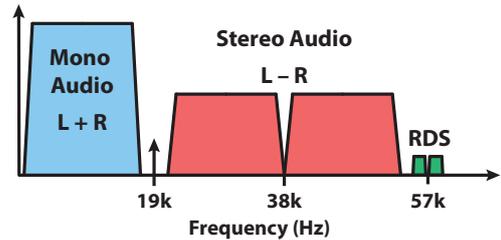

Figure 3: **Structure of FM Radio Transmissions.** Baseband audio signals transmitted in stereo FM broadcasts.

## 3.2 Structure of FM Radio Transmissions

We leverage the structure of FM signals to embed audio and data. In this section, we outline the background about FM radio necessary to understand our design.

*Audio Baseband Encoding.* Fig. 3 illustrates the baseband audio signal transmitted from a typical FM station. The primary component is a mono audio stream, which is an amplitude modulated audio signal between 30 Hz and 15 kHz. With the advent of stereo speakers with separate left and right audio channels, FM stations incorporated a stereo stream transmitted in the 23 to 53 kHz range. To maintain backward compatibility, the mono audio stream encodes the sum of the left and right audio channels (L+R) and the stereo stream encodes their difference (L-R). Mono receivers only decode the mono stream, while stereo receivers process the mono and stereo streams to separate the left (L) and right (R) audio channels. The FM transmitter uses a 19 kHz pilot tone to convey the presence of a stereo stream. Specifically, in the absence of the pilot signal, a stereo receiver would decode the incoming transmission in the mono mode and uses the stereo mode only in the presence of the pilot signal. In addition to the mono and stereo audio streams, the transmitter can also encode the radio broadcast data system (RDS) messages that include program information, time and other data sent at between 56 and 58 kHz.

*RF Encoding.* As the name suggests, FM radio uses changes in frequency to encode data. Unlike packet based radio systems such as Bluetooth and Wi-Fi, analog FM radio transmissions are continuous in nature. An FM radio station can operate on one of the 100 FM channels between 88.1 to 108.1 MHz, each separated by 200 kHz. Specifically, FM radio signals are broadcast at the carrier frequency $f_c$ in one of these FM bands, and information at each time instant is encoded by a deviation from $f_c$.

Mathematically, an FM transmission can be written as,

$$\text{FM}_{RF}(t) = \cos\left(2\pi f_c t + 2\pi \Delta f \int_0^t \text{FM}_{audio}(\tau) d\tau\right) \quad (1)$$

In the above equation, $f_c$ is the carrier frequency of the RF transmission and $\text{FM}_{audio}(\tau)$ is the baseband audio signal shown in Fig. 3. If we normalize $\text{FM}_{audio}(\tau)$ to be between -1 to 1, then $\Delta f$ is the maximum deviation in frequency from the carrier frequency, $f_c$.

Using Carson's rule [28] we can approximate the bandwidth of the above signal as $2(\Delta f + \max(f_{audio}))$, where $\max(f_{audio})$ is the maximum frequency in the baseband audio signal. For FM radio stations broadcasting mono, stereo, and RDS up to 58 kHz, using a maximum $\Delta f$ of 75 kHz [31] results in a bandwidth of 266 kHz.

*FM Decoding.* The FM receiver first demodulates the incoming RF signal to baseband to obtain $\text{FM}_{BB} = \cos\left(2\pi\Delta f \int_0^t \text{FM}_{audio}(\tau)d\tau\right)$. In order to extract the encoded audio information, the receiver performs a derivative operation. Specifically, the derivative $\frac{d}{dt}\text{FM}_{BB}(t)$ converts the frequency changes to amplitude variations:

$$-2\pi\Delta f \text{FM}_{audio}(t) \sin\left(2\pi\Delta f \int_0^t \text{FM}_{audio}(\tau)d\tau\right)$$

Dividing this by the phase shifted baseband FM signal, $\text{FM}_{BB}\left(t+\frac{\pi}{2}\right)$ recovers the original audio signal, $2\pi\Delta f \text{FM}_{audio}(t)$. Note that the amplitude of the decoded baseband audio signal is scaled by the frequency deviation $\Delta f$; larger frequency deviations result in a louder audio signal. In our backscatter prototype, we set this parameter to the maximum allowable value. We also note that while the above description is convenient for understanding, in practice FM receiver circuits implement these decoding steps using phase-locked loop circuits.

## 3.3 Backscattering FM Radio

Backscattering FM radio transmissions and decoding them on mobile devices is challenging because FM receivers only output the decoded audio, while the backscatter operation is performed on the RF signals. To address this, we show that by exploiting the FM signal structure, we can transform backscatter, which performs a multiplication operation in the RF domain, into an addition operation in the audio domain.

To understand this, let us first look at the backscatter operation. Say we have a signal source that transmits a single tone signal, $\cos(2\pi f_c t)$ at a center frequency, $f_c$. If the backscatter switch is controlled with the baseband signal $B(t)$, it generates the signal $B(t)\cos(2\pi f_c t)$ on the wireless medium. Thus, the backscatter operation performs multiplication in the RF domain. So, when we backscatter ambient FM radio signals, $\text{FM}_{RF}(t)$, we generate the backscattered RF signal, $B(t) \times \text{FM}_{RF}(t)$.

Say we pick $B(t)$ as follows:

$$B(t) = \cos\left(2\pi f_{back}t + 2\pi\Delta f \int_0^t \text{FM}_{back}(\tau)d\tau\right) \quad (2)$$

The above signal has the same structure as the FM radio transmissions in Eq. 1, with the exception that it is centered at $f_{back}$ and uses the audio signal $\text{FM}_{back}(\tau)$. The backscattered RF signal $B(t) \times \text{FM}_{RF}(t)$ then becomes:

$$\cos\left(2\pi f_{back}t + 2\pi\Delta f \int_0^t \text{FM}_{back}(\tau)d\tau\right) \times$$
$$\cos\left(2\pi f_c t + 2\pi\Delta f \int_0^t \text{FM}_{audio}(\tau)d\tau\right)$$

The above expression is a product of two cosines, and allows us to apply the following trigonometric identity: $2\cos(A)\cos(B) = \cos(A+B) + \cos(A-B)$. Focusing on just the $\cos(A+B)$ term[2] yields:

$$\cos\left(2\pi\left[(f_c + f_{back})t + \Delta f \int_0^t \text{FM}_{audio}(\tau) + \text{FM}_{back}(\tau)d\tau\right]\right)$$

The key observation is that the above expression is of the same form as an FM radio signal centered at a frequency of $f_c + f_{back}$ with the baseband audio signal $\text{FM}_{audio}(\tau) + \text{FM}_{back}(\tau)$. Said differently, an FM radio tuned to the frequency $f_c + f_{back}$, will output the audio signal $\text{FM}_{audio}(t) + \text{FM}_{back}(t)$. Thus, by picking the appropriate backscatter signal we can use the multiplicative nature of backscatter in the RF domain to produce an addition operation in the received audio signals. We call this *overlay backscatter* because the backscattered information is overlaid on top of existing signals and has the same structure as the underlying data.

Next, we describe how we pick the parameters in Eq. 2.

*1) How do we pick $f_{back}$?* We pick $f_{back}$ such that $f_c + f_{back}$ lies at the center of an FM channel. This allows any FM receiver to tune to the corresponding channel and receive the backscatter-generated signals. In addition, $f_{back}$ should be picked such that the resulting FM channel is unoccupied. We note that due to the bandwidth of FM transmissions, geographically close transmitters are often not assigned to adjacent FM channels [31]. This, along with changes to station ownership and licensing policies over the years, has left various FM channels empty.

Fig. 4a shows the FM band occupancy in five different US cities. We show both the number of licensed stations in a city as well as the number of stations detected in a particular zip code in that city based on public resources available online [13, 7]. This figure shows that in some cities the number of detectable stations is less than the number of assigned channels, as some licensed stations may no longer be operational. Further, in cities like Seattle, there are more detectable bands than assigned licenses as transmissions from neighboring cities may be detected. The key point however is that a large fraction of 100 FM channels are unoccupied and can be used for backscatter.

Finally, to compute the $f_{back}$ required in practice, we measure the frequency separation between each licensed

---
[2]The $\cos(A-B)$ term can be removed using single-sideband modulation as described in [36].

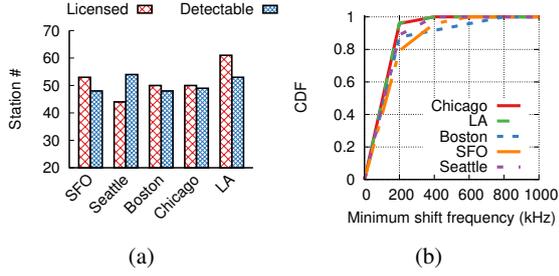
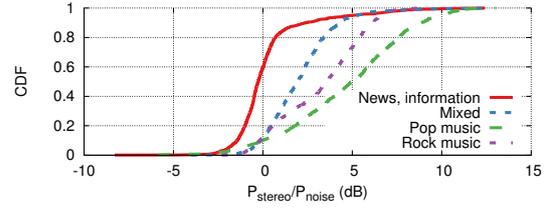

(a)         (b)

Figure 4: **Usage of FM channels in US cities.** (a) shows both the number of licensed and detectable channels of the 100 FM bands and (b) shows the minimum frequency difference between licensed FM stations and the closest unoccupied FM band.

FM station and the nearest channel without a licensed station based on data from [13]. Fig. 4b shows the CDF of the minimum $f_{back}$ across five major US cities. The plot shows that the median frequency shift required is 200 kHz and is less than 800 kHz in the worse case situation. We note that the optimal value of $f_{back}$ will vary by location and should be chosen such that the backscatter transmission is sent at the frequency with the lowest power ambient FM signal. This is because while FM receivers often have very good sensitivities, in practice the noise floor may instead be limited by power leaked from an adjacent channel.

*2) How do we pick $FM_{back}(\tau)$?* This depends on whether we want to overlay audio or data on the ambient FM signals. Specifically, to overlay audio we set $FM_{back}(\tau)$ to follow the structure of the audio baseband signal shown in Fig. 3. To send data, we instead generate audio signals using the modulation techniques described in §3.4.

*3) How do we use backscatter to generate Eq. 2?* At a high level, Eq. 2 is a cosine signal with a time-varying frequency. Thus, if we can generate cosine signals at different frequencies using backscatter, we can create the required backscatter signal. To do this, we approximate the cosine signal with a square wave alternating between +1 and -1. These two discrete values can be created on the backscatter device by modulating the radar cross-section of an antenna to either reflect or absorb the ambient signal at the backscatter device. By changing the frequency of the resulting square wave, we can approximate a cosine signal with the desired time-varying frequencies.

### 3.3.1 FM Backscatter capabilities

The above description focuses on *overlay backscatter* where the backscattered audio data is simply overlaid on top of the ambient FM signals. In this section, we describe two additional backscatter techniques.

**Stereo Backscatter.** We consider two scenarios. 1) A mono radio station that does not transmit a stereo stream, and 2) A stereo station that broadcast news information.

*1) Mono to stereo backscatter.* While many commercial FM radio stations broadcast stereo audio, some stations only broadcast a mono audio stream. In this case, all the frequencies corresponding to the stereo stream (15-58 kHz in Fig. 3) are unoccupied. Thus, they can be used to backscatter audio or data without interference from the audio signals in the ambient FM transmissions. Utilizing these frequencies however presents two technical challenges. First, the FM receiver must be in the stereo mode to decode the stereo stream, and second FM receivers do not provide the stereo stream but instead only output the left and right audio channels (L and R).

To address the first challenge, we note that FM uses the 19 kHz pilot signal shown in Fig. 3 to indicate the presence of a stereo stream. Thus, in addition to backscattering data, we also backscatter a 19 kHz pilot signal. Specifically, our backscatter signal $B(t)$ is:

$$\cos\left(2\pi f_{back}t + 2\pi\Delta f \int_0^t 0.9 FM_{back}^{stereo}(\tau) + 0.1\cos(2\pi 19k\tau)d\tau\right)$$

Inserting this 19 kHz pilot tone indicates that the receiver should decode the full stereo signal which includes our backscattered audio $FM_{back}^{stereo}$.

To address the second challenge, we note that FM receivers do not output the stereo stream (L-R) but instead output the left and right audio streams. To recover our stereo backscatter signal, all we have to do is compute the difference between these left (L) and right (R) audio streams. This allows us to send data/audio in the unoccupied stereo stream of a mono FM transmission.

*2) Stereo backscatter on news stations.* While many FM stations transmit in the stereo mode, i.e. with the 19 kHz pilot tone in Fig. 3, in the case of news and talk radio stations, the energy in the stereo stream is often low. This is because the same human speech signal is played on both the left and right speakers. We verify this empirically by measuring the stereo signal from four different radio stations. We capture the audio signals from these stations for a duration of 24 hrs and compute the average power in the stereo stream and compare it with the average power in 16-18 kHz, which are the empty frequencies in Fig. 3.

Figure 5 shows the CDF of the computed ratios for the four FM stations. These plots confirm that in the case of news and talk radio stations, the stereo channel has very low energy. Based on this observation, we can backscatter data/audio in the stereo stream with significantly less interference from the underlying FM signals. However, since the underlying stereo FM signals already have the 19 kHz pilot signal, we do not backscatter the pilot tone.

**Cooperative backscatter.** Consider a scenario where two users are in the vicinity of a backscattering object, e.g., an advertisement at a bus stop. The phones can share the received FM audio signals through either Wi-Fi direct or Bluetooth and create a MIMO system that can be used to cancel the ambient FM signal and decode the backscattered signal. Specifically, we set the phones to two different FM bands: the original band of the ambient FM signals ($f_c$) and the FM band of the backscattered signals ($f_c + f_{back}$). The audio signals received on the two phones can then be written as,

$$S_{phone1} = FM_{audio}(t)$$
$$S_{phone2} = FM_{audio}(t) + FM_{back}(t)$$

Here we have two equations in two unknowns, $FM_{audio}(t)$ and $FM_{back}(t)$, which we can solve to decode the backscattered audio signal $FM_{back}(t)$. In practice however we need to address two issues: 1) The FM receivers on the two smartphones are not time synchronized, and 2) On the second phone, hardware gain control alters the amplitude of $FM_{audio}(t)$ in the presence of $FM_{back}(t)$.

To address the first issue, we resample the signals on the two phones, in software, by a factor of ten. We then perform cross-correlation between the two resampled signals to achieve time synchronization between the two FM receivers. To address the second issue, we transmit a low power pilot tone at 13 kHz as a preamble that we use to estimate the amplitude. Specifically, we compare the amplitude of this pilot tone during the preamble with the same pilot sent during the audio/data transmission. We scale the power of the received signal by the ratio of the two amplitude values. This allows us to calibrate the amplitude of the backscattered signal. We then subtract these two signals to get the backscatter signal, $FM_{back}(t)$.

### 3.4 Data Encoding with Backscatter

We encode data using the audio frequencies we can transmit using backscatter. The key challenge is to achieve high data rates without a complex modulation scheme to achieve a low power design. High data rate techniques like OFDM have high computational complexity (performing FFTs) as well as have high peak-to-average ratio, which either clips the high amplitude samples, or scales down the signal and as a result limits the communication ranges. Instead we use a form of FSK modulation in combination with a computationally simple frequency division multiplexing algorithm, as described below.

**Data Encoding process.** At a high level the overall data rate of our system depends on both the symbol rate and the number of bits encoded per symbol. We modify these two parameters to achieve three different data rates. We present a low rate scheme for low SNR scenarios as well as a higher rate scheme for scenarios with good SNR.

*100 bps.* We use a simple binary frequency shift keying scheme (2-FSK) where the zero and one bits are represented by the two frequencies, 8 and 12 kHz. Note that both these frequencies are above most human speech frequencies to reduce interference in the case of news and talk radio programs. We use a symbol rate of 100 symbols per second, giving us a bit rate of 100 bps using 2-FSK. We implement a non-coherent FSK receiver which compares the received power on the two frequencies and output the frequency that has the higher power. This eliminates the need for phase and amplitude estimation and makes the design resilient to channel changes.

*1.6 kbps and 3.2 kbps.* To achieve high bit rates, we use a combination of 4-FSK and frequency division multiplexing. Specifically, we use sixteen frequencies between 800 Hz and 12.8 kHz and group them into four consecutive sets. Within each of these sets, we use 4-FSK to transmit two bits. Given that there are a total of four sets, we transmit eight bits per symbol. We note that, within each symbol, there are only four frequencies transmitted amongst the designated 16 to reduce the transmitter complexity. We choose symbol rates of 200 and 400 symbols per second allowing us to achieve data rates of 1.6 and 3.2 kbps. We note that our experiments showed that the BER performance degrades significantly when the symbol rates are above 400 symbols per second. Given this limitation, 3.2 kbps is the maximum data rate we achieve which is sufficient for our applications.

*Maximal-ratio combining.* We consider the original audio from the ambient FM signal to be noise, which we assume is not correlated over time; therefore we can use maximal-ratio combining (MRC) [46] to reduce the bit-error rates. Specifically, we backscatter our data $N$ times and record the raw signals for each transmission. Our receiver then uses the sum of these raw signals in order to decode the data. Because the noise (*i.e.,* the original audio signal) of each transmission are not correlated, the SNR of the sum is therefore up to $N$ times that of a single transmission.

## 4 Implementation

We build a hardware prototype with off-the shelf components, which we use in all our experiments and to build proof of concept prototypes for our applications. We then design an integrated circuit based FM backscatter system and simulate its power consumption.

**Off-the-shelf design.** We use the NI myDAQ as our baseband processor, which outputs an analog audio signal from a file. For our FM modulator, we use the Tektronix 3252 arbitrary waveform generator (AWG) which has a built-in FM modulation function. The AWG can easily operate up to 10s of MHz and can generate an FM modulated square wave as a function of an input signal. Interfacing the NI myDAQ with the AWG gives us the flexibility of using the same setup to evaluate audio and data modulation for both mono and stereo scenarios. We connect the output of the AWG to our RF front end, which consists of the ADG902 RF switch. We design the switch to toggle the antenna between an open and short impedance state.

**IC Design.** In order to realize smart fabric and posters with FM backscatter, we translate our design into an integrated circuit to minimize both size and cost, and scale to large numbers. We implement the FM backscatter design in the TSMC 65 nm LP CMOS process. A detailed description of the IC architecture is presented below.

*Baseband processor.* The baseband data/audio is generated in the digital domain using a digital state machine. We write Verilog code for the baseband and translate it into a transistor level implementation using the Synopsis Design Compile tool. Our implementation of a system transmitting mono audio consumes 1 $\mu$W.

*FM modulator.* The FM modulator is based on an inductor and capacitor (LC) tank oscillator with an NMOS and PMOS cross-coupled transistor topology. We leverage the fact that the capacitors can be easily switched in and out and design a digitally controlled oscillator to modulate the oscillator frequency. We connect a bank of 8 binary weighted capacitors in parallel to an off-chip 1.8 mH inductor and control the digital capacitor bank from the output of the baseband processor to generate a FM modulated output. We simulate the circuit using Cadence Spectre [5]. Our 600 kHz oscillator with a frequency deviation of 75 kHz consumes 9.94 $\mu$W.

*Backscatter switch.* We implement the backscatter switch using an NMOS transistor connected between the antenna and ground terminals. The square wave output of the FM modulator drives the NMOS switch ON and OFF, which toggles the antenna between open and short impedance states to backscatter FM modulated signals. We simulate the switch in Cadence to show it consumes 0.13 $\mu$W of power while operating at 600 kHz. Thus, the total power consumption of our FM backscatter system is 11.07 $\mu$W.

## 5 Evaluation

We evaluate various aspects of our backscatter system.

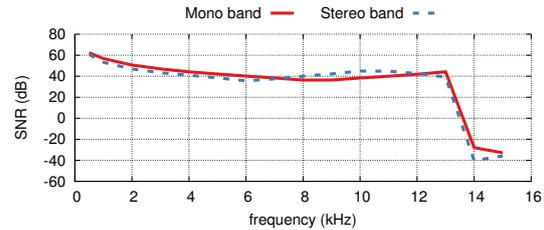

Figure 6: **SNR v/s frequencies.** Received SNR of different frequencies using Moto G1.

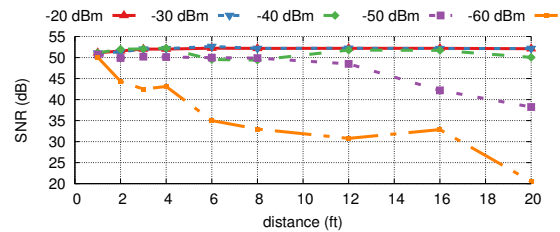

Figure 7: **SNR v/s power and distance.** Received SNR of different receiving powers and distances between the poster antenna and the receiver.

### 5.1 Micro-benchmarks

We first measure the frequency response of FM receivers to understand how they affect audio signals backscattered at different frequencies. We then benchmark the performance of our backscatter system versus distance.

*Characterizing frequency response.* We characterize the frequency response of our whole system by backscattering audio signals at different frequencies and analyzing the received audio signals. In order to benchmark our system without interfering background audio, we simulate an FM station transmitting no audio information (e.g. $FM_{audio} = 0$, which is a single tone signal at the channel center frequency $f_c$) using an SDR (USRP N210). Specifically, we set the USRP to transmit at 91.5 MHz. We then connect our prototype backscatter switch to the poster form factor antenna described in §6.1, which reflects the incoming transmissions and shifts them to an FM channel 600 kHz away by setting $f_{back}$ in Eq. 2 to 600 kHz. We set the backscatter audio $FM_{back}$ to single tone audio signals at frequencies between 500 Hz and 15 kHz.

We use a Moto G1 smartphone to receive these transmissions with a pair of Sennheiser MM30i headphones as its antenna. We use an FM radio app developed by Motorola [11] to decode the audio signals and save them in an AAC format. We perform these experiments in a shielded indoor environment to prevent our generated signals from interfering with existing FM stations.

We separate the backscatter antenna and FM receiver by 4 ft and place them equidistant from the FM transmitter. We then set the power at the FM transmitter such that the received power at the backscatter antenna is -20 dBm. We record the audio signal received on the phone and compute SNR by comparing the power at the frequency corresponding to the transmitted tone and the average power of the other audio frequencies. For instance, if the transmitter sends a single tone at 5 kHz, then we compute the ratio $\frac{P_{5\ kHz}}{\sum_f P_f - P_{5\ kHz}}$. We repeat the same experiment by transmitting the same 500 Hz to 15 kHz tones only in the L-R stereo band to determine whether the stereo receiver affects these signals differently.

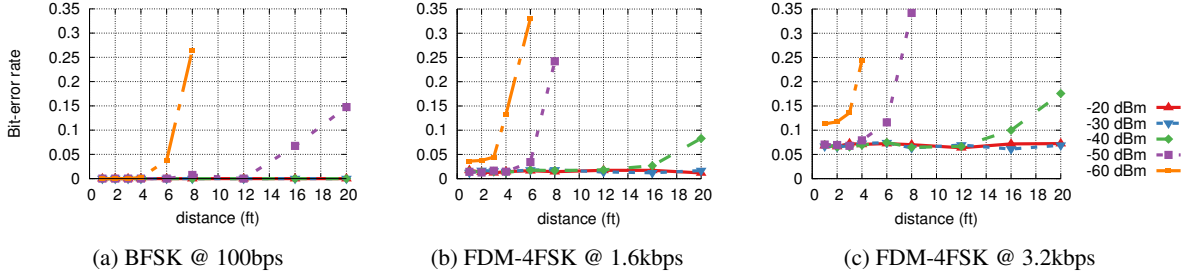

(a) BFSK @ 100bps    (b) FDM-4FSK @ 1.6kbps    (c) FDM-4FSK @ 3.2kbps

Figure 8: **BER w/ overlay backscatter.** Bit-error rates using overlay backscatter *w.r.t.* different receiving powers and distances.

Fig. 6 shows the frequency response for both the mono and stereo streams. The figure shows that the FM receiver has a good response below 13 kHz, after which there is a sharp drop in SNR. While this cut-off frequency is dependent on the FM receiver, recording app, and compression method used to store the audio, these plots demonstrate we can at the least utilize the whole 500 Hz to 13 kHz audio band for our backscatter transmissions.

*Characterizing range versus distance.* In the above set of experiments we fix the distance between the backscatter device and the FM transmitter as well as the power of the FM transmitter. Next, we vary these parameters to understand how the backscatter range changes as a function of the FM radio signal strength. To do this we measure how the SNR changes as a function of the distance between the backscatter device and the FM receiver at five different power levels from -20 dbm to -60 dbm. We set our prototype to backscatter a 1 kHz audio tone using the same setup described above. We then increase the distance between our backscatter device and the receiving phone while maintaining an equal distance from each device to the transmitter.

Fig. 7 shows the results. We can see that the backscatter device can reach 20 ft when the power of the FM transmitter is -30 dBm at the backscatter device. At a -50 dbm power level, the power in the backscattered signal is still reasonably high at close distances. This is because of the good sensitivity of the FM radio receivers in contrast to the TV based approaches explored in prior work such as ambient backscatter [40].

## 5.2 BER performance

Next, we evaluate how well our design can be used to transmit data using FM backscatter. We evaluate the three different bit rates of 100 bps, 1.6 and 3.2 kbps described in §3.4. Our goal in this evaluation is to understand the BER performance of these three different bit rates at different power levels for the ambient RF signals. Since it is difficult to control the power levels of the ambient RF signals in outdoor environments, we create an FM radio station by using a USRP to retransmit audio signals recorded from local FM radio stations. Specifically, we capture 8 s audio clips from four local FM stations broadcasting different content (news, mixed, pop music, rock music) and retransmit this audio in a loop at 91.5 MHz using our USRP setup. This setup allows us to test the effect of different types of background audio in an environment with precisely controllable power settings.

We again set $f_{back}$ for our prototype to be 600 kHz, and set $FM_{back}$ to continuous 8 s data transmissions at each of the bit rates described above. We vary the distance between the backscatter device and the smartphone and adjust the FM transmitter power to control the power received at the backscatter device.

We evaluate the BER with overlay backscatter where the data is embedded in the mono stream on top of the background audio signals transmitted by the USRP. Fig. 8 shows the BER results with the overlaid data as a function of distance between the FM receiver and the backscatter device for the three different bit rates. We also show the results for different power levels at the backscatter device. These plots show that:

• At a bit rate of 100 bps, the BER is nearly zero up to distances of 6 feet across all power levels between -20 and -60 dBm. Further, for power levels greater than -60 dBm, the range increases to over 12 feet. This shows that across all the locations in the survey in §3.1, we can transmit data to the smartphone using FM backscatter.

• Increasing bit rate reduces range. However, at power levels greater than -40 dBm, for both 1.6 and 3.2 kbps, the BERs are low at distances as high as 16 feet. Further, at 1.6 kbps, the BERs are still low up to 3 and 6 feet at -60 and -50 dBm respectively.

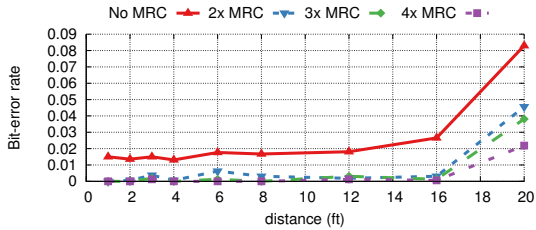

Figure 9: **BER w/ MRC.** Bit-error rates using overlay backscatter with MRC technique.

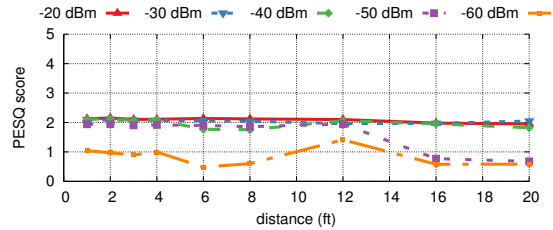

Figure 11: **PESQ w/ overlay backscatter.** PESQ scores of speech received using overlay backscatter *w.r.t.* different receiving powers and distances. In overlay backscatter, we have a background audio from the ambient FM signals and so what we hear is a composite signal. This sounds good at a PESQ value of two.

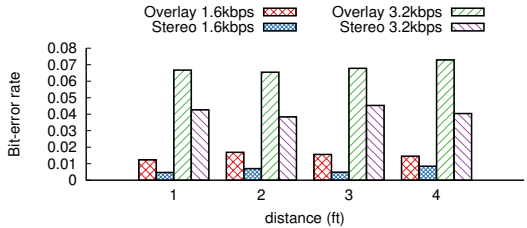

Figure 10: **BER w/ stereo backscatter.** Bit-error rates comparison using overlay backscatter and stereo backscatter.

• We can use the MRC combining technique described in §3.4 to further reduce the BER at the receiver. Fig. 9 shows the BER when performing MRC at a bit rate of 1.6 kbps and power level of -40 dBm. The figure shows the results for combining 2 to 4 consecutive transmissions. The results show that combining across two transmissions is sufficient to significantly reduce BER.

Finally, while MRC reduces the BER it also decreases the effective throughput at the receiver. To prevent this we can use stereo backscatter where the data is encoded in the stereo stream. We repeat a limited set of the above experiments to verify this. Specifically, we set the power received at the backscatter device to -30 dBm and set it to transmit data at bit rates of 1.6 kbps and 3.2 kbps respectively. We use the algorithm described in §3.3 to decode the backscatter information at the FM receiver. Fig. 10 shows the BER achieved using stereo backscatter. For comparison, we also plot the results for overlay backscatter. The plots show that stereo backscatter significantly decreases interference and therefore improves BER at the receiver. While stereo backscatter clearly improves performance, we note that it requires a higher power to detect the 19 kHz pilot signal and can therefore only be used in scenarios with strong ambient FM signals.

### 5.3 Audio Performance

Beyond data, our design also enables us to send arbitrary audio signals using FM backscatter. In this section, we evaluate the performance of FM audio backscatter. As before we set the FM transmitter to send four, 8 s samples of sound recorded from local radio stations. To evaluate the quality of the resulting audio signals, we use perceptual evaluation of speech quality (PESQ) which is a common metric used to measure the quality of audio in telephony systems [35]. PESQ outputs a perception score between 0 and 5, where 5 is excellent quality. We evaluate this metric with overlay, stereo and collaborative backscatter.

*Audio with overlay backscatter.* In the case of overlay backscatter, we have two different audio signals: one from the backscatter device and the second from the underlying FM signals. We compute the PESQ metric for the backscattered audio information and regard the background FM signal as noise. We repeat the same experiments as before where we change the distance between the backscatter device and the Moto G1 smartphone at different power levels of FM signals. We repeat the experiments ten times for each parameter configuration and plot the results in Fig. 11. The plots shows that the PESQ is consistently close to 2 for all power numbers between -20 and -40 dBm at distance of up to 20 feet. We see similar performance at -50 dBm up to 12 feet. Unlike data, audio backscatter requires a higher power of -50 dBm to operate since one can perform modulation and coding to decode bits at a lower data rate at down to -60 dBm, as we showed in Fig. 8. We also note that in traditional audio, a PESQ score of 2 is considered fair to good in the presence of white noise; however our interference is real audio from ambient FM signals. What we hear is a composite signal, in which a listener can easily hear the backscattered audio at a PESQ value of two. We attach samples of overlay backscatter audio signals with PESQ values of 2.5, 2, 1.5 and 1 respectively in the following video for demonstration:
<https://smartcities.cs.washington.edu/pesq.mp4>

*Audio with stereo backscatter.* Next we show that we can backscatter audio information in the stereo stream to minimize the effect of the background FM audio. We run experiments on two different FM signals: a mono only FM station and a stereo news and information station. In the first case, in addition to the audio information, our backscatter device inserts the 19 kHz pilot signal to convert the ambient mono transmission into a stereo

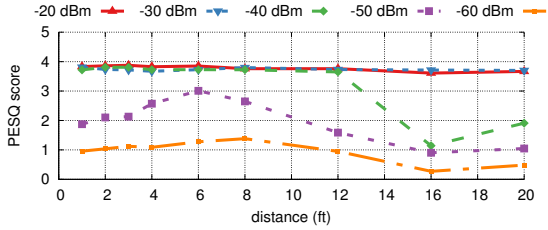

Figure 12: **PESQ w/ cooperative backscatter.** PESQ scores of speech received with cooperative cancellation techniques *w.r.t.* different received powers and distances.

FM signal. We use the same USRP transmitter setup to broadcast an audio signal recorded from a local mono FM station. In both these scenarios, we run the algorithm in §3.3 to cancel the underlying ambient FM audio. We repeat the experiments above and plot the results in Fig. 13a and Fig. 13b. The plots show the following:

• At high FM powers, the PESQ of stereo backscatter is much higher than overlay backscatter. This is because news FM stations underutilize the stereo stream and so the backscattered audio can be decoded with lower noise. At lower power numbers however, FM receivers cannot decode the pilot signal and default back to mono mode. As a result stereo backscatter requires high ambient FM power levels to operate.

• Fig. 13b shows the feasibility of transforming mono FM transmissions into stereo signals. Further, stereo backscatter of a mono FM station results in higher PESQ and can operate at a lower power of -40 dBm. This is because, unlike news stations that have some signal in the stereo stream, mono transmissions have no signals and therefore even less interference than the previous case.

*Audio with cooperative backscatter.* Finally, we evaluate cooperative backscatter using audio transmissions. Specifically, we use two Moto G1 smartphones to achieve a MIMO system and cancel the underlying ambient audio signals. The FM transmitter broadcasts at a frequency of 91.5 MHz and the backscatter device again shifts the transmission by 600 kHz to 92.1 MHz. We set the first smartphone to 92.1 MHz, and the second smartphone to 91.5 MHz and keep them equidistant from our backscatter prototype. We use the cancellation algorithm in §3.3 to decode the backscattered audio signals. We repeat the above set of experiments and plot the results in Fig. 12. The plots show that cooperative backscatter has high PESQ values of around 4 for different power values between -20 and -50 dBm. This is expected because the MIMO algorithm in §3.3 cancels the underlying audio signal. We also note that cooperative backscatter can operate with much weaker ambient FM signals (-50 dBm) than stereo backscatter (-40 dBm). This is because as explained earlier, radios do not operate in the stereo mode when the incoming FM signal is weak.

### 5.4 Using FM Receivers in Cars

We evaluate our backscatter system with an FM radio receiver built into a car in order to further demonstrate the potential for these techniques to enable connected city applications. The FM receivers built into cars have two distinct differences compared to smartphones. First, car antennas can be better optimized compared to phones as they have less space constraints, the body of the car can provide a large ground plane, and the placement and orientation of the antenna can be precisely defined and fixed unlike the loose wires used for headphone antennas. Because of this, we expect the RF performance of the car's antenna and radio receiver to be significantly better than the average smartphone. Second, although recent car models have begun to offer software such as Android Auto or Apple CarPlay, the vast majority of car stereos are still not programmable and therefore limited to using our overlay backscatter technique.

To test the performance of the car receiver with our backscatter system, we use a similar experimental setup to the one described above for the smartphone. Specifically, we place the backscatter antenna 12 ft away from the transmitting antenna which we configure to output a measured power, and evaluate the quality of the audio signal received in a 2010 Honda CRV versus range. Because the radio built into the car does not provide a direct audio output, we use a microphone to record the sound played by the car's speakers. To simulate a realistic use case we perform all experiments with the car's engines running and the windows closed. We backscatter the same signals used above to measure SNR and PESQ. Figure 14 shows the audio quality versus range for two different power values, demonstrating our system works well up to 60 ft.

## 6 Proof-of-concept Applications

We evaluate two proof-of-concept applications to show how FM backscatter applies to real world scenarios.

### 6.1 Talking Posters

Posters and billboards have long been a staple of advertising in public spaces. However these traditional methods of advertising rely entirely on a single graphic to attract attention and convey information. A poster capable of broadcasting to a smartphone could send audio, notifications and links for discounted event tickets. This added level of interactivity allows the advertisement to persist beyond a quick glance and allows the advertiser to take advantage of smartphone functionalities, for example to give directions to an event.

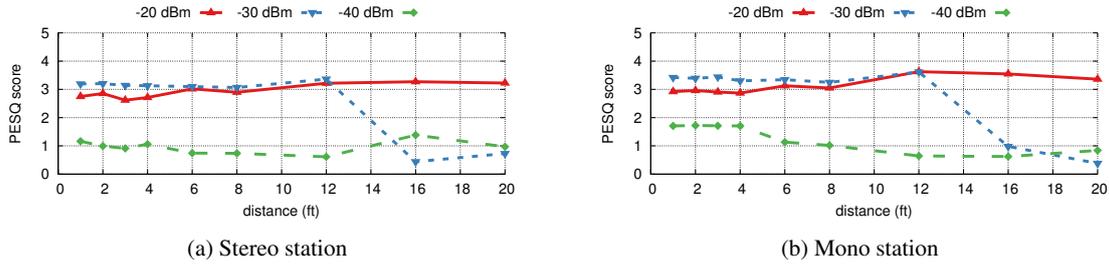

(a) Stereo station  (b) Mono station

Figure 13: **PESQ w/ stereo backscatter.** PESQ scores of speech received with stereo backscatter techniques *w.r.t.* different receiving powers and distances: (a) is for sending audio in the stereo stream of a stereo broadcast and (b) is for transforming a mono station into a stereo transmission.

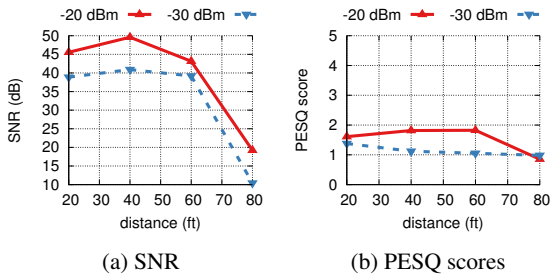

(a) SNR  (b) PESQ scores

Figure 14: **Overlay backscatter performance in a car.** SNR and PESQ scores recorded in a car /wrt different receiving power and distances.

We leverage the low power backscatter techniques described in this paper to show that posters can broadcast information directly to cars and smartphones in outdoor environments. To evaluate this, we design two poster form factor antennas. We fabricate a half wavelength dipole antenna on a bus stop size poster (40"x60"), as well as a bowtie antenna in the common Super A1 poster size (24"x36"). Both are fabricated by placing copper tape (Tapes Master) in the desired pattern onto a sheet of 45 lb poster paper. Fig. 15a shows the 24"x36" prototype.

To evaluate this system, we place our prototype poster antenna on the side of a real bus stop, as shown in Fig. 15b. We backscatter ambient radio signals from a local public radio station broadcasting news and information at a frequency of 94.9 MHz. At the poster location we measure an ambient signal power of -35 dBm to -40 dBm. We test both data and audio transmissions using overlay backscatter. We create our backscatter signal at 95.3 MHz. A user standing next to the poster wearing a pair of headphones connected to a Moto G2 smartphone records these backscattered signals. Our results show that we can decode data at 100 bps at distances of up to ten feet. Fig. 16 shows an advertisement for the band *Simply Three* as a notification on a smartphone. We also overlay a snippet from the band's music on top of the ambient news signals and record the audio decoded by the phone at a distance of 4 ft from the poster.

In addition to evaluating the poster with smartphones, we also evaluate this system using the FM radio receiver built into a car to further demonstrate how these poster and sign form factor devices could be used for connected city applications. We mount the poster on a wall 5 ft above the ground and park a 2010 Honda CRV 10 ft from the poster. To simulate a realistic use case in an urban environment, we position the poster on the side of the building without direct line of sight to the FM transmitter. We again backscatter the ambient radio signal available at 94.9 MHz to 95.3 MHz and use a microphone to record the audio played from the car's speakers. We attach sample clips of the above recordings at the following link:
<https://smartcities.cs.washington.edu/apps.mp4>

## 6.2 Smart Fabric

There has been recent interest in smart fabrics that integrate computing, sensing and interaction capabilities into garments. Given their unique advantage of being in contact with the body, textiles are well suited for sensing vital signs including heart and breathing rate. In fact, clothing manufacturers have begun to explore integrating sensors into different types of clothing[22, 3]. More recently, Project Jacquard from Google [44] designs touch-sensitive textiles to enable novel interaction capabilities. We explore the potential of using FM radio backscatter to enable connectivity for smart fabrics. Low power connectivity can enable smart fabrics that do not require large batteries, which is useful for reducing form factor and improving durability.

We design a prototype of our backscatter system on a shirt as shown in Fig. 17a. Specifically, we use Ansys HFSS to design and simulate a meander dipole antenna small enough to fit on the front of a standard 15 in wide adult small T-shirt. Building on the antenna designs in [33, 26], we fabricate our prototype on a 100% cotton t-shirt by machine sewing patterns of a conductive 3 ply thread made out of 316L stainless steel fibers [17] which does not oxidize even after repeated use or washing.

Wearable systems suffer from losses such poor antenna performance in close proximity to the human body and must perform consistently while a user is in motion. To evaluate whether our backscatter system is suitable for these applications, we connect our backscatter switch to the prototype shirt antenna and use it to transmit data at both 100 bps and 1.6 kbps. We perform this experiment in an outdoor environment in which the prototype antenna receives ambient radio signals at a level of -35 dBm to -40 dbm. Fig. 17b compares the bit error rate when the user is standing still, running (2.2 m/s) or walking (1 m/s). The plot shows that at a bit rate of 1.6 kbps while using MRC, the BER was roughly 0.02 while standing and increases with motion. However at a lower bit rate of 100 bps, the BER was less than 0.005 even when the user was running. This demonstrates that FM radio backscatter can be used for smart fabric applications.

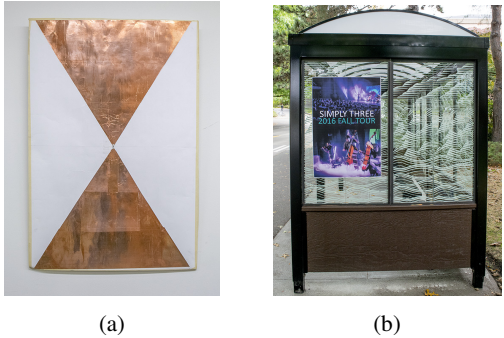

Figure 15: **Talking Poster application.** The figures show both a close up of our poster form factor antenna as well as its deployment at a bus stop.

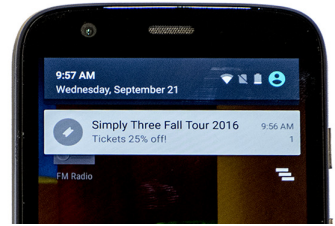

Figure 16: Example of a notification sent by a poster advertising discounted tickets to a local concert.

## 7 Related Work

Our work is related to RFID systems [32, 10, 48, 51, 34, 41] that use expensive readers as dedicated signal sources. The challenge in using RFID systems outdoors is the cost of deploying and maintaining the RFID reader infrastructure. This has in general tempered the adoption of RFID systems in applications beyond inventory tracking.

Ambient backscatter [40, 42] enables two RF-powered devices to communicate with each other by scattering ambient TV signals. In contrast, our focus is to transmit to mobile devices such as smartphones using backscatter communication. Since smartphones do not have TV receivers, TV signals are unsuitable for our purposes. Furthermore, given the transition to digital TV, the number of available broadcast TV channels has been declining over the years [27]. So it is unlikely that future smartphones will incorporate TV receivers.

[37] backscatters transmissions from a Wi-Fi router, which can be decoded on existing Wi-Fi chipsets using changes to the per-packet CSI/RSSI values. BackFi [43] improved the rate of this communication with a full-duplex radio at the router to cancel the high-power Wi-Fi transmissions from the reader and decode the weak backscattered signal. Passive Wi-Fi [38] and [29] demonstrated the feasibility of generating 802.11b and Bluetooth transmissions using backscatter communication. Both these systems require infrastructure support in terms of a plugged in device that transmits a single tone signal. Given the scarcity of Wi-Fi outdoors as well as the challenges in deploying plugged-in devices, these approaches are not applicable in outdoor environments.

More recently, work on FS-backscatter [49] and Interscatter [36] demonstrate that one can backscatter either Bluetooth or Wi-Fi transmissions from a device (e.g., smartphone or router) and decode the backscattered signals on another device (e.g., smart watch). These approaches require two different devices to be carried by the user, which is not convenient. Further, given the lukewarm adoption of smart watches [21], it is important to explore the solution space for techniques that can work without them. In contrast, this paper shows that one can leverage the ambient FM signals in outdoor environments as a signal source to enable transmissions from everyday objects to mobile devices such as smartphones.

Bluetooth [4], NFC [50] and QR-codes [39] are alternate technologies to enable connectivity. Our backscatter design is orders of magnitude lower power and cheaper (5-10 cents [9]) than Bluetooth since it does not require generating an RF signal. Further, it does not require pointing the phone camera toward the QR-codes and has a better range than NFC, which is limited to a few cm. Additionally, the ranges we achieve allow multiple people to easily view and interact with the poster at once unlike NFC or QR codes.

Finally, FCC regulations allowing weak unlicensed transmitters in FM bands [30] led to the proliferation of personal FM transmitters [19, 20]. These transmitters however consume orders of magnitude more power and are therefore not suitable for our target applications.

## 8 Discussion and Conclusion

This paper opens a new direction for backscatter research by showing for the first time that ambient FM radio signals can be used as signal sources for backscatter communication. In this section, we discuss various aspects of this technology and outline avenues for future research.

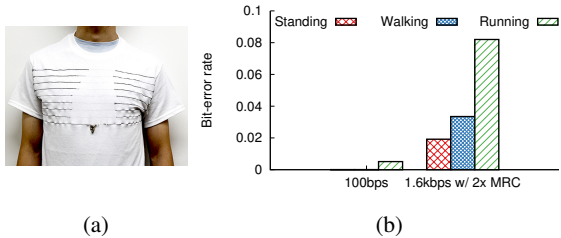

(a) (b)

Figure 17: **Smart fabric application.** (a) shows a cotton shirt with an antenna sewn onto the front with conductive thread and (b) shows BER for the shirt in various mobility scenarios.

*Multiple backscatter devices.* Since the range is currently limited to 4–60 feet, multiple devices that are spread out can concurrently operate without interference. For backscattering devices that are closer to each other, one can set $f_{back}$ to different values so that the backscattered signals lie in different unused FM bands. We can also use MAC protocols similar to the Aloha protocol [25] to enable multiple devices to share the same FM band.

*Improving data ranges and capabilities.* We can use coding [42] to improve the FM backscatter range. We can also make the data transmission embedded in the underlying FM signals to be inaudible from a user perspective by using recent techniques for imperceptible data transmission in audible audio frequencies [47].

*Power harvesting.* We can explore powering these devices by harvesting from ambient RF signals such as FM or TV [42, 45] or using solar energy that is often plentiful in outdoor environments. We note that the power requirements could further be reduced by duty cycling transmissions. For example, a poster with an integrated motion sensor could transmit only when a person approaches.

*Potpourri.* One can also explore dual antenna designs that use Wi-Fi as an RF source indoors and switch to FM signals in outdoor environments. To enable the link from the phone to the backscatter device, we can use the ultra low-power Wi-Fi ON-OFF keying approach described in [36]. Finally, many phones use wired headphones as FM radio antennas. Although Apple recently removed the standard 3.5 mm headphone connector from its latest iPhone model, Apple offers an adapter to connect wired headphones [2]. Thus, future adaptors could be designed for use with headphones as FM antennas.

## 9 Acknowledgments.

We thank Deepak Ganesan for his helpful feedback on the paper. This work was funded in part by awards from the National Science Foundation (CNS–1452494, CNS–1407583, CNS–1305072) and Google Faculty Research Awards.